# Request for Comments: Proposal of a Blockchain for the Automatic Management and Acceptance of Student Achievements

Thorsten Sommer[1], Gergana Deppe[1], Valerie Stehling[1], Max Haberstroh[1], Frank Hees[1]
[1]Cybernetics Lab IMA & IfU, RWTH Aachen, Germany

E-Mail: thorsten.sommer@rwth-aachen.de, gergana.deppe@ima-ifu.rwth-aachen.de, valerie.stehling@ima-ifu.rwth-aachen.de, max.haberstroh@ima-ifu.rwth-aachen.de, frank.hees@ima-ifu.rwth-aachen.de

**Introduction**

Staying abroad during their studies is increasingly popular for students: In 2012, 4.5 million students went abroad (2.3%), and a number of up to 6.4 million (3.2%) is predicted for 2025 [2]. However, there are various challenges for both students and universities. One important question for students is whether or not achievements performed at different universities can be taken into account for either enrolling at a foreign university or for completing the studies at their home university [3]. This is not only important for students who want to spend a semester abroad, but also for those who change universities or study courses within a university or decide to enroll for a second course of study. Within Europe, the Bologna process aimed to make academic achievements compatible between European countries [4]. In the meantime, all study programs have been converted to the Bachelor and Master system, making study achievements comparable by means of e.g. credit points.

However, the process of comparing and approving the academic performance of students still needs a lot of manual labor. The question arises of how a harmonization or at least the coordination of academic achievements can be realized and automated worldwide, cf. [3]. For example, an international masters' degree program on engineering entrepreneurship is currently being developed in the research project ELLI, Excellent Teaching and Learning in Engineering Sciences. It is intended that at least three universities on three continents will implement this course of study together.

In addition to university achievements, an increasing proportion of the 195 million students worldwide [1] increasingly receive certificates from "massive open online courses" – MOOCs – or other social media services [5]. The integration of such services into university teaching is still in the initial stages and presents some challenges [6]. In the meantime, higher education policy is currently being overtaken by the labor market: An account at the social media service Github, the largest collaborative platform for open source developments, is now an essential prerequisite for employers [7]. Furthermore, alternatives to the university system are already emerging, where commercial providers offer their own study programs based on publicly available MOOCs [8].

In this paper we describe the idea to manage all these study achievements worldwide in a decentralized system, which might solve the national and international challenges regarding the recognition of student achievements. On the student side, the motivation is based on less bureaucracy and at the same time more freedom in the choice of courses (both subject-related and geographic). On the organization side, the long-term motivation is to reduce the administrative effort and thus lower costs while at the same time speeding up processing.

The aim of this paper is to encourage discussion in the global community instead of presenting a finished concept. This discussion is important for a holistic debate and critical questioning of the open issues. In addition to the technical aspects, the organizational aspects in particular must be discussed, as the idea presented here might involve far-reaching changes.

**Concept of Blockchains**

In 2008 the concept of the blockchain was presented [9]. In the following, the principle is briefly explained: A blockchain is a decentralized construct, which therefore does not require central control and regulation and even prevents centralized intervention [10]. Technically, a blockchain is not implemented using clients and servers, but instead via peer-to-peer networks in which all peers have equal rights [9]. Transactions, data and contracts – the so-called smart contracts – can be stored in the "blocks" of the blockchain. The integrity of the blockchain and the data stored in it is ensured by means of cryptography. A blockchain is protected against manipulation because it can be viewed publicly, and a separate copy is maintained on each peer. A malicious change by a peer is immediately noticeable, as all other copies would be different. This is an important feature to ensure data provenance. In addition, each block is secured with a signature so that the authenticity of the data record can be verified along the complete chain. For a more in-depth description of the blockchain concept, cf. [10]–[12].

Blockchains have become known through so-called crypto currencies [13], e.g. Bitcoin [9] and Ethereum [14]. In the case of crypto currencies, financial transactions are stored in the blocks of the chain. These global systems do not require banks or similar trusted third parties as intermediaries. In addition to data, a block can also contain a smart contract [15]. These contracts fulfill themselves, automatically, if the previously defined conditions for their fulfillment have been met by the data stored in the chain.

**Concept of a Blockchain for Students' Achievements**

In this section, the concept of block chains is applied to students' achievements. First, the network architecture is discussed, followed by an approach to the write permission of the chain. Subsequently, a description of how achievements might be stored in the blocks as well as a description of the smart contracts is following.

Blockchains based on smart contracts are particularly suitable for the homogeneous recognition of student achievements of any institutions. Therefore, we propose to set up a public blockchain for such achievements. The peers can be operated by any person or organization without compromising security. The scalability of such a blockchain is given by the peer-to-peer architecture. This eliminates the need to run powerful servers at corresponding costs. The necessary computing power is distributed across all peers. However, a blockchain can become very large [16]. Usually, each peer has a complete replica of the chain, resulting in an unacceptable demand on mass storage. For this reason, so-called light clients have already been introduced for Ethereum, which work without a local copy of the chain [17]. In this case, the concept of equal peers is diminished in favor of practicality.

In addition to universities, other organizations must also be able to write e.g. open badges of the Mozilla foundation [18] or other types of certificates in the blockchain. An approach to prevent users from writing useless entries might look like this: As usual in blockchains, a user has to solve some kind of cryptographic challenge to write. Over time, these challenges become harder to solve, so the computational effort increases. This measure is called proof-of-work [19], [20]. If the tasks become too computationally complex, it is possible to switch to another method such as proof-of-sake [21] at a later point in time. As a second measure, anyone wishing to write in the blockchain might be required to register and document their identity publicly in the chain. It should be possible to read the blockchain publicly via a web interface, whereby the data is only stored anonymously. Two types of data are stored in the blocks of the chain: The achievements of students and smart contracts.

If students have taken an e-exam, their results are automatically saved in the blockchain. In order to guarantee the data protection of the students, they should receive a randomly generated number of which identity is not publicly known. A so-called random "Universal Unique Identifier", UUIDv4, would fit well, since these are not predictable and collision free [22]. In addition to the result of the assessment, metadata of the lecture should also be included, such as the number of credit points, the time required, the issuing organization, and a list of topics covered. This metadata is required as the basis for the smart contracts. In the event of a posterior error in the documentation of achievements, the issuing organization must write a new block in the chain to correct the existing block. Direct correction of the incorrect entry is not possible due to the design of blockchains.

As soon as two organizations e.g. universities or MOOC providers agree on the acceptance of student achievements, they can document this agreement in the form of a smart contract. In addition, a university or faculty can define a smart contract for a course of study by documenting the achievements to be made for the degree. These smart contracts fulfil themselves as soon as a student has provided all the necessary achievements. The difference to the agreements between organizations known today is the language of smart contracts. Instead of natural language, which allows different formulations and ambiguities, formalized languages, similar to programming languages, are used. Their syntax ensures that the definition and thus also the interpretation of a smart contract is unambiguous.

**Use Cases**

Figure 1 shows various use cases, which are explained below. The cases chosen are limited to universities, however, they are examples and do not exclude e.g. providers of MOOCs, etc. As can be seen in the figure in the fictional blocks 101 and 102, the students' achievements are persisted in independent blocks. In the case of electronic assessments, the result can theoretically be published automatically in the chain. In the case of manual pen and paper assessments, the examination office must publish the results after manual correction and marking. The advantage for students is that they can take an assessment at any university in the world without any further formalities. It is sufficient to show the blockchain ID and an official identification document to ensure identity. This approach would significantly reduce the bureaucratic obstacles to a stay abroad, change the university or switch the course of study.

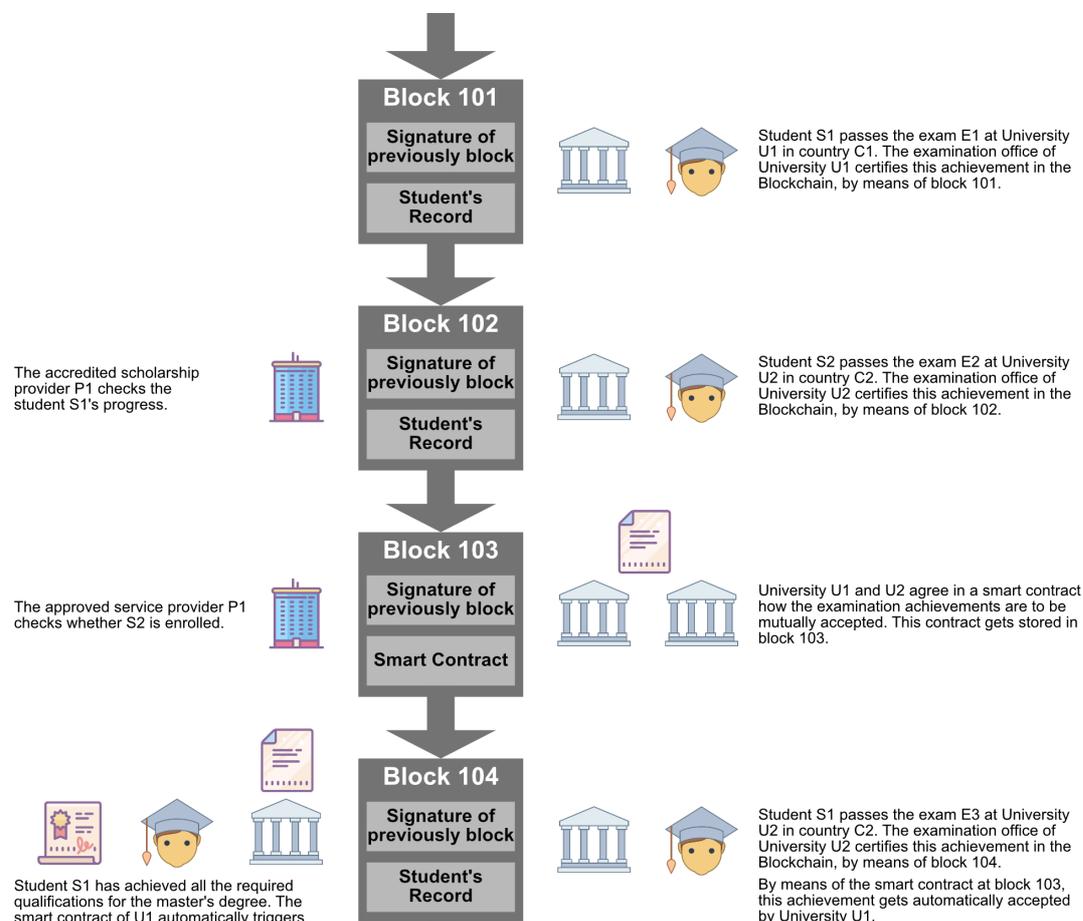

*Figure 1: Overview of the simplified blockchain concept. The icons are provided by Icons8*

Another use case is a query on the progress of studies from accredited scholarship providers. These can be government agencies or private institutions. In this way, the responsible authorities can check the conditions of a scholarship, so that the bureaucratic effort is also minimized by this second use case. Figure 1 shows two further use cases by means of block 103 and 104. As soon as a student performs an exam abroad as an example, this is automatically recognized by the home university. This is done by applying the relevant smart contract, e.g. of block 103. Also for this step the student does not have to make an application or take any other steps.

Finally, the last use case shows that the graduation can also run completely automatically. The matching smart contract from the home university recognizes that all achievements have been made. The certificate can be printed, signed, dispatched and the exmatriculation can be initiated.

**Further Research**

We are aware that this idea has far-reaching consequences for (university) education worldwide. Primal, the idea of a worldwide blockchain for student achievement must be discussed and refined before it can be implemented. A quick implementation is neither useful nor appropriate, as various challenges remain unresolved.

First and foremost, the issue of data protection must be discussed in the community. Is it sufficient if every student receives an unpredictable ID? Failed assessments and multiple assessments by students are publicly documented, even if this information is anonymous. Deleting a record in the blockchain is technically impossible. Errors on the part of an examination office can only be corrected by a new block. The old incorrect entry is still documented. This is also both an advantage and a disadvantage. It is conceivable that the data of the blocks are additionally encrypted and the student is the only one who knows the key. This would ensure data protection. For a smart contract to become effective, the student would have to decrypt the blocks on his computer with his own data. If the contract can be fulfilled, a new block with the result would be created, encrypted and written into the chain.

Secondly, there is the question of the value of university degrees. Do bachelor's and master's degrees become less valuable if students can listen to the necessary lectures at any institution? Is it advantageous to blend lectures at universities with MOOCs and other achievements or does this dilute the educational system? This question might be of particular interest to higher education policy.

Thirdly, cases of deliberate counterfeiting must be discussed. As soon as an entry in the blockchain is recognized as illegal, a new block must be created for correction, which declares the older entry as invalid. As a consequence of such an event, the issuing organization would have to be sanctioned if the event is not an isolated incident. One possible sanction might be that in the event of repeated incidents, the right to write archivements in the blockchain is withdrawn. For this case, a smart contract could be defined that automatically searches for correction blocks or withdrawal blocks and determines the frequency per organization. If a threshold value is exceeded, this contract prevents this organization from continuing to write data.

Fourthly, the blockchain idea described here would open up new possibilities for optimization. Interested students could, for example, train an artificial intelligence to find the simplest or cheapest combination for a Master's degree. Is this kind of optimization an argument against the development of such a blockchain? Are such activities insignificant for the education system?

Such a blockchain would be technically feasible with foreseeable effort, since known blockchains have often been implemented as open source, which can be used to build on previous experience. However, it is necessary to discuss the idea holistically beforehand, to develop an appropriate concept and to conduct the necessary political debate in order to influence higher education policy. An immediate implementation would result in an empty shell, since the data (student achievements) and the smart contracts might be missing. Thus, higher education policy would have to openly commit to such a system, otherwise it would be meaningless to implement it. A blockchain only becomes fully effective if it is used worldwide. This also applies to this idea: ultimately, a global agreement would be necessary to give students the freedoms set out in this idea.

**Conclusion**

In this paper an idea was presented, which brings together the technical innovation of a blockchain with the requirements of organizations, e.g. universities, regarding students' archivements and the need of students for less bureaucracy. After the description of the problems of the students and the institutes, a short introduction to the topic blockchain followed. Besides the concept of a blockchain for exam results of students, various use cases were discussed. Finally, further research needs were identified.

The idea presented would remove bureaucratic borders between universities and the countries of the world for 195 million students. Students could attend lectures and take exams at any university at any time, without uncertainty about the recognition of achievements. The concept described would also enable a revival of a Studium generale, if this is desired by universities. This paper is intended to provide the momentum for this concept to be discussed in the community.

**Acknowledgment**

This work is part of the project ELLI, "Excellent Teaching and Learning in Engineering Sciences," and was funded by the federal ministry of education and research ("BMBF"), Germany.

**Keywords**

Blockchain, Internationalization, Social Media, MOOCs, E-Assessments